\documentclass[preprint,aps,nofootinbib,tightenlines]{revtex4}
\usepackage{latexsym}
\usepackage{amssymb}
\usepackage{amsmath}
\usepackage[hypertex]{hyperref}
\usepackage{graphicx}

\begin{document}

\preprint{hep-ph/0411133}

\vspace*{1.5cm} 
\title{ Dark Matter from Baryon Asymmetry  \vspace{0.5cm}}

\author{Ryuichiro Kitano and Ian Low}
\affiliation{School of Natural Sciences, Institute for Advanced Study, 
           Princeton, NJ 08540
\vspace*{1cm}
}

\begin{abstract}
\vspace*{1cm}

The measured densities of dark and baryonic matter are surprisingly
close to each other, even though the baryon asymmetry and the dark
matter are usually explained by unrelated mechanisms. We consider a
scenario where the dark matter $S$ is produced non-thermally from the
decay of a messenger particle $X$, which carries the baryon number and
compensates for the baryon asymmetry in the Universe, thereby
establishing a connection between the baryonic and dark matter
densities. We propose a simple model to realize this scenario, adding
only a light singlet fermion $S$ and a colored particle $X$ which could 
have a
mass in the ${\cal O}$(TeV) range and a lifetime to appear long-lived in
collider detector.
Therefore in hadron colliders the signal is similar to that
of a stable or long-lived gluino in supersymmetric models.

\end{abstract}











\maketitle

\section{Introduction}

Our current understanding of the evolution of the Universe is based on
the standard cosmology, the Friedman-Robertson-Walker cosmological
model, augmented by the standard model of particle physics. It is
clear that this understanding is, however, by no means complete,
especially after the tremendous increase in both the volume and
accuracy of data from cosmological observations. The standard model,
while passed very stringent tests from particle physics experiments
and by itself is capable of describing fundamental interactions in
energies all the way up to the Planck scale, comes out short in
several ways when trying to explain the cosmological data. The data
now support, for examples, a dark energy responsible for the
accelerating expansion of the Cosmos, a solid case for the non-baryonic
dark matter, a nearly scale-invariant, adiabatic, and Gaussian density
fluctuations favored by inflation, and a baryon-asymmetric Universe,
all of which cannot be accommodated by the standard model alone.

Brought upon us by the observations,
these new insights in turn provide many
directions to ponder physics beyond the standard model. In most
scenarios, different shortcomings of the standard model are rectified
by different and unrelated mechanisms, which makes it a wonder when
some observed values from seemingly different physical origins are
close to each other. An example is the ratio of the baryon and dark
matter densities. Sakharov \cite{Sakharov:1967dj} pointed out that
baryogenesis can
be achieved by three ingredients: baryon number violation, C and CP
violation, and a departure from thermal equilibrium. On the other
hand, dark matter is usually proposed to explain the observed galaxy
rotation curves, distributions and clustering of galaxies,
gravitational lensing effects, the power spectrum of the cosmic
microwave background, and so on \cite{Eidelman:2004wy}, none of which
seem directly relevant to the baryon asymmetry. Nevertheless, the
measured dark matter density turns out to be quite close to the
baryonic matter density \cite{Eidelman:2004wy,Spergel:2003cb}:
$\Omega_{\rm DM}/\Omega_b \sim 5$.

Therefore it is natural to look for models in which these two densities
have a common origin, and thereby explaining the proximity of the two
numbers. 
In this case, the mass of the dark-matter particle is predicted 
from the ratio of $\Omega_{\rm DM}/\Omega_b$ 
to be several GeV if the two number densities are comparable. This is an
interesting value, being not far from the electroweak scale at
which we suspect new physics may appear.
There have been a number of attempts in this regard over the years
\cite{Barr:1990ca,Barr:1991qn,Kaplan:1991ah,Dodelson:1991iv,
Kuzmin:1996he,Farrar:2004qy,Hooper:2004dc}.
The central idea remains that the dark-matter abundance comes from
an asymmetry in some new quantum number generated at the same time as
the baryogenesis, thus providing the link between the baryon density, or more
accurately the nucleonic density we observed, and the dark matter
density.
Also note that there are proposals based on ideas from different
directions~\cite{Thomas:1995ze,Fujii:2002aj,Ibe:2004yp,Catena:2004pz,Wilczek:2004cr,Allahverdi:2004ds}.

Here we consider a simple mechanism discussed in
Ref.~\cite{Kuzmin:1996he}, where dark matter is remnant of an asymmetry
of a quantum number in a separate sector, in which all the particles are
charged under a new symmetry while the standard model particles are
neutral.
The lightest particle charged under this new symmetry is stable and a
natural dark matter candidate. At the time of baryogenesis, the $B-L$
number is split between the standard model and the dark sector, the
sector charged under the new symmetry. If subsequently the interactions
between the standard model and the dark sector are switched off or
negligible, the $B-L$ number is separately conserved in the two sectors.
The excess of the $B-L$ number in the standard model, which results in
the baryon asymmetry in the Universe, is then compensated by the dark
sector, establishing the link between the dark matter and baryonic
matter densities. Such an idea, when applies to the minimal supersymmetric
standard model \cite{Kuzmin:1996he}, suffers from various phenomenological
difficulties; among others, the asymmetries in both sectors will be washed 
out by scattering processes through the gaugino-exchange diagrams. 
%

Generally speaking, in order for such an idea to work, 
the dark matter candidate needs to have a large enough annihilation
cross section so that the dark matter number density, which is the
sum of the dark matter candidate and its anti-particle, can be linked
to the baryon number density, which is proportional to the difference
in the dark matter candidate and its anti-particle.
Since we need the dark matter to be neutral, this 
implies the annihilation process has to rely on either the $Z$ boson
exchange or extra gauge/Yukawa interactions. For the $Z$ exchange, it
is severely constrained by the invisible $Z$ decay width as well as
the direct detection searches. For extra gauge/Yukawa interactions, obviously
additional model buildings are needed which are likely to be
complicated.

In this article, we adopt a minimalistic approach, in a similar fashion
as in \cite{Davoudiasl:2004be}, and propose a very simple yet realistic
model in which dark matter is produced {\it non-thermally} by the
late-time decay of heavy particles. In addition to the standard model,
we introduce the smallest possible new symmetry, a $Z_2$ symmetry, along
with two new particles $S$ and $X$, both of which are odd under the
$Z_2$ symmetry and comprise the dark sector.
The dark matter $S$ is a gauge-singlet fermion who interacts with the
standard-model sector through the messenger particle $X$.
By the interaction between the standard model and the messenger
particle, the baryon (or $B-L$) number is split between the two
sectors. The asymmetry in the messenger particle is then converted into
the dark matter by the non-thermal decay into $S$, giving a dark matter
number density similar to that of the baryon.
Such a simple model turns out to have interesting collider phenomenology, 
which will be discussed later.

In the next section, we discuss the scenario mentioned above in general
terms, followed by a section considering the cosmological constraints on
various aspects of the scenario. After that we explicitly write down a
simple model realizing the scenario by introducing the dark matter as a
gauge singlet fermion and the messenger as a heavy, colored
particle. Then we study the collider phenomenology of our simple
model. The lifetime of the messenger particle can be as long as
$10^{-2}$ sec, resulting in collider signals very similar to that of a
long-lived gluino in split supersymmetry \cite{Arkani-Hamed:2004fb} in
the Large Hadron Collider (LHC). In the end we summarize and conclude.

\section{The General Scenario}

The basic ingredients of the scenario are simply the standard model plus
the dark matter $S$ and the messenger particle $X$, as well as a new
$Z_2$ symmetry which we call $T$-parity.\footnote{Incidentally, or
perhaps not so much so, the little hierarchy problem strongly suggests a
new symmetry at the TeV scale in order to stabilize the electroweak
scale naturally, which can be just a $Z_2$ \cite{Cheng:2003ju}.}  All
the standard model particles have even $T$-parity, whereas $S$ and $X$
are odd. The fermion $S$ is an electroweak singlet and the lightest
$T$-odd particle (LTP), but we do not specify the quantum numbers of $X$
here since there are a wide range of possibilities. Then the scenario
proceeds in three stages as follows.

During the first stage,
the baryogenesis is made possible without $B-L$ violation by 
distributing the $B-L$ number between the standard model ($T$-even)
and the dark sector ($T$-odd). A simple, but not unique, way to achieve 
this is through the out-of-equilibrium and CP violating decay of a heavy
$T$-odd particle $P$ into a quark and the messenger particle $X$. 
An asymmetry in $B-L$ is generated in each sector even though
the net $B-L$ number is vanishing. (However the heavy 
particle $P$ is not necessary; see later section for an alternative
scenario without having to introduce it.)
Furthermore, we assume that the interactions between the standard model and the
dark sector are decoupled below the temperature of the baryogenesis.
Thus effectively we have two separately conserved $B-L$ numbers in the two
sectors. There may be interactions which preserve both the $B-L$
symmetry and the $T$-parity such as $X \bar{X} H H^\dagger$, where $\bar{X}$
refers to the anti-particle of $X$. 
Such interactions, however, do not re-distribute the $B-L$ numbers. As
such, $T$-parity and gauge symmetry can guarantee the $B-L$ numbers
to be conserved separately by appropriately choosing the quantum
number of $X$, which result in the following relation:
\begin{eqnarray}
 n_{B-L}^{\rm SM} = - n_{B-L}^{X} = - q_{B-L} ( n_{X} - n_{\bar{X}} ) \ ,
\label{bmnumber}
\end{eqnarray}
where $q_{B-L}$ is the $B-L$ charge of the messenger $X$, and
$n_{B-L}^{\rm SM}$ and $n_{B-L}^{X}$ are the $B-L$ number densities in the
standard model and the dark sector, respectively. 

On the other hand, since both $X$ and $\bar{X}$ eventually decay into the LTP, 
the dark matter candidate $S$, its number density is given by the total
number of $X$ and $\bar{X}$ particles
\begin{equation}
n_{\rm DM} = n_X^{\rm tot} \equiv n_X + n_{\bar{X}},
\end{equation}
which is independent of the $n_{B-L}^{\rm SM}$ in Eq.~(\ref{bmnumber})
and would suggest there is {\it no} connection between the baryonic and
dark matter densities, unless $n_X \gg n_{\bar{X}} \sim 0$ or the other
way around. This implies the lifetime of $X$ should be long enough so
that it does not decay until after most of the $\bar{X}$ particles
annihilate with $X$. Therefore, the second stage of the scenario is the
annihilation of the messenger particle in the dark sector. At
temperature $T < m_X$, where $m_X$ is the mass of the messenger,
particle $X$ starts to annihilate with its anti-particle $\bar{X}$
through gauge interactions and we are left with an abundance of
$X$. Consequently,
\begin{eqnarray}
 n_{B-L}^{\rm SM} = - n_{B-L}^{X} \simeq - q_{B-L}\  n_{X}^{\rm tot}\ .
\end{eqnarray}

The final step is the decay of $X$ into the dark matter $S$. 
We emphasized that, in order to establish the link between the number densities
of the dark and
baryonic matter, the messenger $X$ needs to have a lifetime long 
enough to survive until
after the annihilation is completed. This will be the case if there is 
no relevant or marginal operator contributing to the decay of $X$, which 
can be achieved easily by choosing the spin and/or the quantum number of $X$.
Then the $X$ decay produces
the same number density of the LTP as $n_X^{\rm tot}$ due to $T$-parity.
At this point there are two distinct situations. One is that $X$ decays after the
electroweak phase transition. The baryon and the dark matter number
densities in this case are given by
\begin{eqnarray}
 n_B = \left( \epsilon - \frac{q_{B}^{\rm decay}}{q_{B-L}} \right)
n_{B-L}^{\rm SM} \ , \ \ \ 
 n_{\rm DM} = \left| \frac{n_{B-L}^{\rm SM}}{q_{B-L}} \right|
 \ ,
\label{eq:B-and-DM}
\end{eqnarray}
where $q_{B}^{\rm decay}$ is the effective baryon number of $X$ defined
by the operator which induces the $X$ decay. The efficiency $\epsilon$
is the relation between the $B-L$ number and the baryon asymmetry in the
presence of the sphaleron process. This may be different from the
standard model value of 28/79 \cite{Harvey:1990qw} since the $U(1)_Y$
neutrality condition is modified by the asymmetry stored in $X$ which
may have a non-vanishing $U(1)_Y$ charge.
The second term in the baryon number density is the contribution from
the decay products of the $X$ particle.
The other possibility is that $X$ decays before the electroweak phase
transition. We obtain a different formula from Eq.~(\ref{eq:B-and-DM}):
\begin{eqnarray}
 n_B = \epsilon \left( 1 - \frac{q_{B-L}^{\rm decay}}{q_{B-L}} \right)
n_{B-L}^{\rm SM} \ , \ \ \ 
 n_{\rm DM} = \left| \frac{n_{B-L}^{\rm SM}}{q_{B-L}} \right|
 \ ,
\label{eq:B-and-DM-2}
\end{eqnarray}
where $q_{B-L}^{\rm decay}$ is the effective $B-L$ charge of $X$ defined
by the operator which induces the $X$ decay. In general, it is different
from $q_{B-L}$ which is the charge defined by the operator responsible
for the baryogenesis. 
We will see an example of such a case later, even though the decay in 
the example occurs
after the electroweak symmetry breaking owing to the cosmological 
constraints to be discussed in the next section.

 From Eqs.~(\ref{eq:B-and-DM}) and (\ref{eq:B-and-DM-2}) and the observed ratio
 of $\Omega_{\rm DM}/\Omega_b$, we can determine the mass of the
 dark matter $m_{\rm DM}$ from $m_p$, the mass of the proton:
 \begin{eqnarray}
 \frac{m_{\rm DM}}{m_p} 
= \frac{\Omega_{\rm DM}}{\Omega_{B}} 
\left| \frac{n_B}{ n_{\rm DM}} \right|
= 5.1 \left|
 \epsilon\, q_{B-L} - q_B^{\rm decay}
\right| \ ,
\label{eq:relation}
\end{eqnarray}
for $X$ decaying after the electroweak phase transition and 
\begin{eqnarray}
 \frac{m_{\rm DM}}{m_p} 
= 5.1\, \epsilon\, \left|
q_{B-L} - q_{B-L}^{\rm decay}
\right| \ ,
\label{eq:relation2}
\end{eqnarray}
for $X$ decaying before the electroweak phase
transition. 
The mass of the dark matter $S$ is predicted to be of ${\cal O}$(GeV) if
its number density is comparable to that of the baryon.

\section{Cosmological Constraints}

There are only two model parameters in this scenario after fixing $m_{\rm DM}$
to give the correct ratio of $\Omega_{\rm DM}/\Omega_b$. One is
the scale $M$ that suppresses the operator for the decay of $X$, while
the other is the mass of the messenger particle $m_X$.  Several
constraints on these two parameters arise from cosmological
observations. The first requirement, if $X$ decays after the electroweak
phase transition, is the decay should happen before the big-bang
nucleosynthesis, $\tau \lesssim 10^{-2}$ s, in order not to tamper with this very
stringent test of the standard cosmology.
Suppose $X$ decays through a dimension $D$ operator with $D>4$, ${\cal
O}_{\rm decay}= {\cal O}_{\rm SM} (X S) / M^{D-4}$, an upper bound on $M$ 
is given by
the following
\begin{eqnarray}
 M \lesssim 10^{15} {\rm ~GeV}
\left(
\frac{m_X}{1 {\rm ~TeV}}
\right)^{\frac{3}{2}} \ ,
\label{eq:bound-on-M5}
\end{eqnarray}
for $D=5$ and 
\begin{eqnarray}
 M \lesssim 10^9 {\rm ~GeV}
\left(
\frac{m_X}{1 {\rm ~TeV}}
\right)^{\frac{5}{4}} \ ,
\label{eq:bound-on-M6}
\end{eqnarray}
for $D=6$.
If instead we require the decay to occur before the electroweak phase transition,
\begin{eqnarray}
 M \lesssim 10^{11} {\rm ~GeV}
\left(
\frac{m_X}{1 {\rm ~TeV}}
\right)^{\frac{3}{2}} \ ,
\end{eqnarray}
for $D=5$ and 
\begin{eqnarray}
 M \lesssim 10^7 {\rm ~GeV}
\left(
\frac{m_X}{1 {\rm ~TeV}}
\right)^{\frac{5}{4}} \ ,
\end{eqnarray}
for $D=6$.

On the other hand, a lower bound on $M$ can be obtained 
from the requirement that the
decay of $X$ should occur after the completion of the annihilation, which
results in
\begin{eqnarray}
 M \agt  m_X \left(  \frac{M_{\rm Pl}\ x_f^2}{m_X}
\right)^{\frac{1}{2D-8}} \ ,
\label{mlower}
\end{eqnarray}
where $x_f$ is defined as $x_f \equiv m_X / T_f$, with $T_f$ being the
freeze-out temperature of the annihilation, and is evaluated to be 
${\cal O}(20)$ \cite{Kolb:1990vq}.

As for the mass of the messenger particle $X$, it is constrained by the
requirement that there are sufficient annihilations so that the
symmetric component of $X$ (the total abundance minus asymmetry) becomes
much less significant than asymmetry. 
%
Within the standard model gauge interactions, the cross section from the
strong interaction is estimated to be $0.2/m_X^2$ for
$s$-wave annihilation which gives an upper bound on $m_X$:
\begin{eqnarray}
 m_X \ll 2 \times 10^4 {\rm ~TeV} \left( \frac{5 {\rm ~GeV}}{m_S} \right)  \ .
\label{bound-on-mX}
\end{eqnarray}
For the case where $X$ is a scalar particle, the annihilation through
the gluon-exchange diagram is $p$-wave, which gives a stronger bound
than above:
\begin{equation}
m_X \ll  8 \times 10^2 {\rm ~TeV} \left( \frac{5 {\rm ~GeV}}{m_S} \right) \ .
\end{equation}
These bounds are larger than the unitarity bound of 350 TeV \cite{Griest:1989wd} 
derived for a stable massive particle which was once in thermal equilibrium. This
is because our messenger particle eventually decays into the dark matter $S$ which
is much lighter than $X$, resulting in a much smaller mass density. Hence $m_X$ can
be much larger than 350 TeV without overclosing the Universe.
We stress that the values given here
are simply upper bounds on the mass of the messenger, which are
saturated when the symmetric component of the messenger $X$ is equal to the 
asymmetric component. In order to establish the connection between the baryon and
dark matter number densities, $n_{B-L}^{\rm SM} \sim -q_{B-L} n_X^{\rm tot}$, 
it is preferred to have $m_X$ much lower than
the upper bounds. On the other hand, the lower bounds on $m_X$ simply come from 
direct searches of new particles which is less than 500 GeV. Therefore it is quite
natural for the messenger particle to have a mass in
the ${\cal O}$(TeV) range, which raises the interesting possibility that 
it could be observed at future collider experiments.

There is also a constraint on the reheating temperature of the Universe
by the requirement that the number of thermally produced $S$ particles through
${\cal O}_{\rm decay}$ must be smaller than that of the non-thermal component 
from the $X$ decay. The bound is given by
\begin{eqnarray}
 T_{\rm RH} \lesssim M
\left(
 10^{-7} \frac{M}{M_{\rm Pl}} 
\right)^{\frac{1}{2 D - 9}}\ .
\label{eq:bound-on-Trh}
\end{eqnarray}
For baryogenesis to work, we need at least $T_{RH} > m_X$ to generate
asymmetry in the $X$ particle. This gives a constraint on $M$ 
 more stringent than the lower bound derived from 
Eq.~(\ref{mlower}):
\begin{eqnarray}
 M \gtrsim 10^{14} {\rm ~GeV} 
\left(
\frac{m_X}{1 {\rm ~TeV}}
\right)^{\frac{1}{2}}\ ,
\label{eq:M_from_reheating_5}
\end{eqnarray}
for $D=5$ and 
\begin{eqnarray}
 M \gtrsim 10^{8} {\rm ~GeV} 
\left(
\frac{m_X}{1 {\rm ~TeV}}
\right)^{\frac{3}{4}}\ ,
\label{eq:M_from_reheating_6}
\end{eqnarray}
for $D=6$.
Note that, however, if $X$ decays through a dimension six operator, there
is a dimension five operator, $(SS)(H^\dagger H)/M_{S}$, which could contribute 
to the thermal production of $S$ dominantly unless $M_{S}$ satisfies the lower
bound Eq.~(\ref{eq:M_from_reheating_5}), which we assume to be the case.

By comparing Eqs.~(\ref{eq:M_from_reheating_5}) and (\ref{eq:M_from_reheating_6}) 
with the mass
bound in Eq.(\ref{bound-on-mX}), we conclude that the decay always
occurs after the electroweak phase transition, and our scenario is very predictive 
on the mass scale $M$:
\begin{equation}
\label{Mconstraint1-summary}
10^{14} {\rm ~GeV} \left(\frac{m_X}{\rm 1~TeV} \right)^\frac12 
\lesssim M \lesssim
10^{15} {\rm ~GeV}  \left(\frac{m_X}{\rm 1~TeV} \right)^\frac32, 
\end{equation}
for $D=5$ and 
\begin{equation}
\label{Mconstraint2-summary}
10^{8} {\rm ~GeV} \left(\frac{m_X}{\rm 1~TeV} \right)^\frac34 
\lesssim M \lesssim
10^{9} {\rm ~GeV}  \left(\frac{m_X}{\rm 1~TeV} \right)^\frac54, 
\end{equation}
for $D=6$.
The lifetime of $X$, for $m_X = 1$~TeV, ranges from $10^{-5}$
($10^{-7}$) to $10^{-2}$ second for $D=5$ ($D=6$). For comparison, in
the LHC a particle with a lifetime longer than $10^{-6}$ second will
decay outside of the detector and appear to be stable.
%

\section{A Simple Model}

In this section we present an explicit model and discuss its collider
phenomenology in the following section. We will take the quantum numbers
of $X$ to be the same as the anti-particle of the right-handed down type
quark $X:({\bf \bar{3}}, {\bf 1})_{1/3}$ with spin $1/2$.  Baryogenesis
can be achieved by the out-of-equilibrium and CP-violating decay of a
singlet $T$-odd particle $P$, which can be the inflaton, into $d_R+X$
and $\bar{d}_R + \bar{X}$.
The effective $B-L$ number of $X$ is then $q_{B-L} = -1/3$.
The annihilation of $X$ is sufficiently effective because of the strong
interaction once the bound in Eq.(\ref{bound-on-mX}) is satisfied.

With this assignment of quantum numbers and spin for $X$, gauge symmetry and
$T$-parity ensures the lowest dimensional operator contributing to the $X$ decay
is a dimension six one:
\begin{eqnarray}
 {\cal O}_{\rm decay} = \frac{1}{M^2} u^c d^c X S \ .
\end{eqnarray}
Thus the bounds on the scale $M$ are given by Eq.~(\ref{Mconstraint2-summary}).
The decay operator breaks the $B-L$ symmetry and defines $B$ and $B-L$ 
numbers of $X$ to  be different from
$q_{B-L}$,
\begin{eqnarray}
 q_{B}^{\rm decay} = q_{B-L}^{\rm decay} = +\frac{2}{3}\ .
\end{eqnarray}
We then obtain the number densities $n_B$ and $n_{\rm DM}$ by applying
Eq.~(\ref{eq:B-and-DM}) as follows:
\begin{eqnarray}
 n_B = ( \epsilon + 2 ) n_{B-L}^{\rm SM} \ ,\ \ \ 
 n_{\rm DM} = 3 | n_{B-L}^{\rm SM}|\ ,
\end{eqnarray}
where the efficiency $\epsilon$ is different from the standard model
value, now that the presence of $X$ modifies the charge neutrality
condition, and calculated to be $34/79$ with the additional constraint
that the total $(B-L)$ number is zero before the decay of $X$. Therefore
the mass ratio $m_{\rm DM} / m_p$ is given by
\begin{eqnarray}
 \frac{m_{\rm DM}}{m_p} = 4.1 \ ,
\end{eqnarray}
from Eq.~(\ref{eq:relation}). The observed ratio of $\Omega_{\rm DM}/\Omega_b$
determines the mass of the dark matter to be 3.9~GeV in this model.

Alternatively, if one were to insist on not introducing the extra heavy
particle $P$, a possibility would be that the baryogenesis generates
$B-L$ asymmetry (or asymmetry in $X$) first and distribute the asymmetry
to the $T$-odd ($T$-even) sector through an interaction such as $(d_R
X)(d_R X)$.
After the decoupling of the interaction, the $B-L$ asymmetry conserves
separately in each sector and the asymmetry in $X$ again becomes the
source of the dark matter.  The prediction to the ratio $m_{\rm DM} /
m_p$ is modified to be
\begin{eqnarray}
 \frac{m_{\rm DM}}{m_p} = 0.89 \ .
\end{eqnarray}

We mention in passing that models in which $X$ decays through
dimension-five operators can also be easily constructed with $X$ being a
scalar particle. For example, $X$ can again have the quantum numbers
$X:({\bf \bar{3}}, {\bf 1})_{1/3}$ but with spin 0 this time. It decays
through the dimension five operator ${\cal O}_{5}=(\bar{q}_L H)^* (X
S)/M$ with $M$ satisfying the constraint
Eq.~(\ref{Mconstraint1-summary}). Note that there is actually a
dimension four operator ${\cal O}_4=d^c \bar{X} S$ allowed by the gauge
symmetry and the $T$-parity. However, we can impose a Peccei-Quinn
symmetry or its discrete subgroup, under which the singlet fermion $S$
is charged, to allow ${\cal O}_{5}$ but prohibit ${\cal O}_4$ at tree
level. Then the Peccei-Quinn symmetry is only softly broken by the mass
term of $S$, which would radiatively induce the operator ${\cal O}_4$
with a small coefficient $m_S/(16\pi^2 M)$. Then $X$ still decays
dominantly through ${\cal O}_{5}$ since $\Gamma_{{\cal
O}_4}/\Gamma_{{\cal O}_5} \sim (m_S/16\pi^2 m_X)^2$ which is only
$10^{-10}$ for $m_S$= 1 GeV and $m_X$=1 TeV. In the end the mass ratio
$m_{\rm DM}/m_p$ is 0.97 or 6.0 for the case with or without the $P$
particle.
Or more simply, if one did not insist on standard model quantum numbers,
dimension four operators could be forbidden by appropriately choosing
quantum number of $X$.
%

\section{collider signals}

In spite of the simplicity of this model, the collider signal for the
messenger particle turns out to be quite interesting. As discussed before,
it is  natural for the messenger particle $X$ to
have a mass in the ${\cal O}$(TeV) range and a lifetime between $10^{-7}
s$ to $10^{-2} s$, which implies its collider phenomenology shares
similar features with that of a heavy stable/long-lived
particle. Examples of such particles include a heavy gluino as the
lightest supersymmetric particle (LSP) \cite{Baer:1998pg} and a
long-lived gluino in the scenario of split supersymmetry
\cite{Arkani-Hamed:2004fb}. The phenomenology of such a
long-lived/stable gluino has been studied extensively in
Refs.~\cite{Raby:1998xr,Mafi:1999dg,Kilian:2004uj,Hewett:2004nw,Cheung:2004ad,Arkani-Hamed:2004yi}. Here
we very briefly summarize their results and point out differences, if
any.

At the LHC $X$ can be pair-produced from $s$- and $t$-channels through
$gg$ and $\bar{q}q$ annihilations if it is colored. Because of its long
lifetime, $X$ hadronizes into a color-singlet state before decaying. In
supersymmetry such particles, resulting from the hadronization of the
gluino, are called $R$-hadrons since they carry one unit of
$R$-parity. In our case, the role of $R$-parity is played by the
$T$-parity so we may as well call the corresponding color singlets
$T$-hadrons. In the model discussed in the previous section, unlike
gluino, $X$ is a fermion in the fundamental representation of $SU(3)_c$
with the quantum numbers of the $b$-quark, which implies the
spectroscopy of $T$-hadrons should look very much like that of hadrons
containing a $b$-quark. That is, there should be states like $Xq$, a
spin 0 $T$-{\it meson}, and $X\bar{q}\bar{q}$, a spin 1/2 $T$-{\it
baryon}. It is well-known that, in the case of $b$-quark, heavy quark
effective theory (HQET) \cite{Manohar:2000dt} is a useful tool in
studying the spectroscopy and interactions of $B$-meson. Since the
messenger particle $X$ is one thousand times heavier than the $b$-quark,
we expect that HQET will be extremely useful in studying the
spectroscopy and interactions of the $T$-hadron since all the $1/m_X$
corrections are very small.

In terms of detection at the LHC, with a lifetime longer than
$10^{-7}$~s, $X$ is most likely to decay outside of the detector and
appears to be a massive stable particle as far as the collider is
concerned. Then the technique for searching for the gluino LSP should be
employed.
In the scenario where the messenger $X$ hadronizes into neutral
particles, interacts very softly with the detector, and remain so
through out their lifetime, pair-productions of $X$ in the collider will
become invisible to the detector, and the detection will rely on
pair-production of $X$ plus an additional jet for the event to be
triggered as the monojet with missing energies carried away by the
neutral $T$-hadrons \cite{Hewett:2004nw}.
An event easier to observe is a pair-production of $X$ hadronizing into
charged $T$-hadrons, which can be distinguished from light particles by
looking at their velocities and the energy loss due to ionization
\cite{Hewett:2004nw}. In the LHC we expect that the $T$-hadron can be
discovered for $m_X \lesssim 2$~TeV, a bound similar to that of a
long-lived gluino \cite{Kilian:2004uj}.

There is also an interesting possibility that $X$ and $\bar{X}$ would
form a bound state, a quarkonium $\eta_4$, in the hadron
colliders~\cite{Barger:1987xg}. For discovery potential at the LHC,
studies found that the most promising channel is $\eta_4 \to \gamma
\gamma$ \cite{Arik:2002nd}. For luminosity at 100~fb$^{-1}$, $\eta_4$
can be observed up to $m_{\eta_4} = 400$ GeV.

\section{Summary}

The origins of baryon asymmetry and dark matter are two cosmological puzzles 
for which the standard model of particle physics fails to explain; the 
CP violation in the
standard model is too small to explain the baryon asymmetry and there is simply
no candidate for dark matter in the particle content of standard model.
Conventionally the dark matter is postulated to be a weakly interacting 
particle whose relic density is determined by the freezing out of the 
interactions that keep it in thermal equilibrium, and is independent of the
dynamics generating the baryon asymmetry. Therefore it is a mystery that
the observed values of the dark and baryonic matter densities are quite close to
each other, within a factor of 5.

In this article we study a general scenario where the dark matter is
produced non-thermally from the decay of a messenger particle, which
carries the $B-L$ number and compensates for the excess of the baryon
number in the standard model. The simplest realization of this scenario
includes only two new particles, the dark matter $S$ which is a gauge
singlet fermion with a mass in the ${\cal O}$(GeV) range and the
messenger particle $X$ which is a colored particle with possibly a mass in the
${\cal O}$(TeV) range. Moreover, the $X$ particle has a long lifetime,
between $10^{-7}$~s and $10^{-2}$~s, which makes interesting collider
signals and, to the first order approximation, mimics that of a
stable/long-lived gluino in supersymmetric models with a gluino LSP or
split supersymmetry.

Finally, it seems appropriate to try to implement this idea in more
complicated models for physics beyond standard model. Earlier attempts
in this regard typically involve the thermal production of sneutrino as
the dark matter in supersymmetric versions of the standard
model~\cite{Kuzmin:1996he,Hooper:2004dc}, and efforts need to be made to
make the sneutrino dark matter much lighter, or its number density much
larger, than natural expectations. In this regard the scenario proposed
in \cite{Agashe:2004ci,Agashe:2004bm}, where a dark matter candidate with a baryon
number and a mass as low as ${\cal O}$(GeV), seems promising.

\acknowledgements

We thank K.~Agashe for bringing Ref.~\cite{Agashe:2004ci} to our
attention. We also acknowledge helpful conversations with T.~Watari 
on the bounds on the mass of the messenger particle. This work is supported by 
funds from the Institute for
Advanced Study and in part by the DOE grant number DE-FG02-90ER40542.


\end{document}